\def\bea{\begin{eqnarray}}
\def\eea{\end{eqnarray}}

\def\beq{\begin{equation}}
\def\eeq{\end{equation}}
\def\ba{\beq\begin{array}{c}}
\def\ea{\end{array}\eeq}

  \newcommand{\bn}{\begin{enumerate}}
\newcommand{\en}{\end{enumerate}}
\newcommand{\bi}{\begin{itemize}}
\newcommand{\ei}{\end{itemize}}

\def\tilde{\widetilde}

\def\hat{\widehat}

\def\cech{${\rm C}^{\kern-6pt \vbox{\hbox{$\scriptscriptstyle\vee$}\kern2.5pt}}${\rm ech}}
\def\Cech{${\sl C}^{\kern-6pt \vbox{\hbox{$\scriptscriptstyle\vee$}\kern2.5pt}}${\sl ech}}

\def\CD{{\cal D}}

\def\CO{{\cal O}}

\def\CS{{\cal S}}

\def\dd{{\bf p}}

     \def\CO{{\cal O}}

\def\uu{ u}

\def\inv{^{-1}}

 \def\frac#1#2{{\textstyle{#1\over#2}}}
\def\inv{^{\raise.15ex\hbox{${\scriptscriptstyle -}$}\kern-.05em 1}}
 \def\hf{\frac{1}{2}}

\def\({\left(}
\def\){\right)}
\def\<{\left\langle\,}
\def\>{\, \right\rangle}
\def\[{\left[}
\def\]{\right]}

\newcommand{\no}{\nonumber}

\def\la{\label}

\documentclass[12pt,a4paper]{article}
\input epsf
\usepackage{amsmath}
\usepackage{amssymb}
\usepackage{graphicx}
\usepackage{mathrsfs}
\usepackage{ulem}

\newcommand{ \Li}{{\rm Li_2}}
\newcommand{ \BDS}{{\rm BDS}}
\newcommand{ \BES}{{\rm BES}}
\newcommand{\bpi}{ \mbox{\boldmath${\pi}$}}

\begin{document}
\title{A note on the  eigenvectors of long-range spin chains and their scalar products}
\author{Didina Serban,\\ {\it {\small Institut de Physique Th\'eorique, DSM, CEA, URA2306 CNRS,}}\\ {\it {\small Saclay, F-91191 Gif-sur-Yvette, France}}}

\maketitle

\abstract{ In this note, we propose an expression for the eigenvectors and scalar products for a class of spin chains with long-range interaction and $su(2)$ 
symmetry.  This class includes the Inozemtsev spin chain as well as the BDS spin chain, which is a reduction of the one-dimensional Hubbard model at half-filling
 to the spin sector. The proposal is valid  for large spin chains and is based on the construction of the monodromy matrix using the Dunkl operators. For the 
 Inozemtsev model these operators are known explicitly. This construction gives in particular the eigenvectors of (an operator closely related to) the dilatation operator of the ${\cal N}=4$ gauge theory in the $su(2)$ sector up to three-loop order, as well as their  scalar products. We suggest how this will affect  the  
expression for the quasi classical limit of the three-point functions obtained by I. Kostov and how to include the all-loop interaction. } 
\bigskip

{\bf Introduction:} Integrability is a powerful tool in theoretical physics.  The most spectacular recent achievement it underlies  is to determine  the spectrum of 
the  
 ${\cal N}=4$ SYM theory for an arbitrary coupling constant and to make very precise quantitative predictions confirming the validity the AdS/CFT correspondence 
 \cite{AdSCFT}. For recent reviews on this subject see \cite{habilitation,Beisert-Rev}.
 The correlation functions \cite{RV,Tailor1, Tailor23, Orselli, GV}, the amplitudes \cite{aTBA} and the Wilson loops \cite{wlTBA, Janik} seem also to be within the reach of 
 integrability methods. At weak coupling, the problem of computing the three-point function of single-trace operators at tree level in the $su(2)$ sector was 
 reformulated \cite{Okuyama:2004bd,RV,Tailor1} in the language of spin chains in terms of scalar products the XXX spin chain.
The scalar products are expressible as determinants \cite{Izergin, Korepin, Slavnov}. The determinantal formula initially proposed for single-chain scalar products was generalized \cite{Foda} to the 
overlaps involving several spin chains. 
Several special cases were considered \cite{Tailor1, Tailor23} involving at least one operator which is close to BPS. The main difficulty in taking the classical 
limit, which is necessary to compare with the string predictions, resides in taking continuum limit of large Slavnov determinants.  This difficulty was overcome 
very recently by I. Kostov \cite{Ivan3pt}, who succeeded to obtain a factorization formula for the Slavnov determinants. He obtained the 
continuum limit of the correlator of three operators with three sets of $su(2)$ charges.
At higher loop order, the dilatation operator is mapped to a spin chain with long range interaction \cite {BKS}. In \cite{SS}, it was matched  with the Inozemtsev 
model \cite{ino02} up to three loop order. Another related chain which reproduces the dilatation operator up to three loops is the BDS spin chain \cite{BDS} which can be be obtained by reduction \cite{RSS} of the one-dimensional Hubbard 
model at half-filling to the spin sector.
Both these models belong to a class of deformations of the XXX model considered in \cite{Boost}. These models were considered out of reach of the algebraic 
Bethe ansatz techniques, which allows to build the eigenvectors of the of the Hamiltonian, because the monodromy matrix was not known.  The sole exception was the Inozemtsev model at infinite length \cite{SS}, whose monodromy matrix can be obtained by analogy with the one for the Haldane-Shastry model \cite{Polychronakos:1992ki,BGHP,HT}. The construction is based on an (iso)morphism with the inhomogeneous XXX model.
Here, we work 
out the details for the Inozemtsev spin chain and we postulate that an identical construction underlies all the models which are constructed from the boost charges 
in \cite{Boost}, including the BDS model.  The Hamiltonian obtained from the monodromy matrix at finite length has a periodicity defect but, on the basis of the perturbative results obtained by Gromov and Vieira for the periodic finite chains \cite{GV}, we think that it is possible to modify the construction to include perfectly periodic chains. It is not clear whether the lack of periodicity of the Hamiltonian really is a drawback in the context of  the computations of the correlation functions in the ${\cal N}=4$ SYM theory, since 
the procedure of splitting and joining of the chains effectively breaks the periodicity. We certainly need to understand the origin of the cancellation between the contact terms necessary to restore the periodicity and the Hamiltonian insertions at the splitting/joining points which was observed at one loop in \cite{GV}.
 This work was triggered by discussions with N.~Gromov
and P.~Vieira and by their recent results \cite{GV} on the eigenvectors of the dilatation operator and the scalar products at two and three loops.

{\bf The Inozemtsev spin chain:} The Inozemtsev spin chain \cite{ino02} is an integrable long-range interacting spin chain with Hamiltonian
\begin{equation}
\label{ino}
H=\sum_{j=1}^L  \sum_{n=1} ^{L-1} {\cal P}_{L,\pi/\kappa}(n) (1-P_{j,j+n})\;,
\end{equation}
where ${\cal P}_{L,\pi/\kappa}(z)$ is the Weierstrass function with periods $L$ and $i\pi/\kappa$ and $P_{j, j+n}$ permutes the spins at sites $j,j+n$.  In \cite {SS} it has been shown that around 
the limit $\kappa \to \infty$ this Hamiltonian 
matches the dilatation operator of the ${\cal N}=4$ gauge theory in the $su(2)$ sector up to three-loop order. In fact, the dilatation operator is a combination
of the Hamiltonian (\ref{ino}) and a higher conserved quantity, but this aspect is irrelevant to the construction of the eigenvectors.
The full solution of the Hamiltonian (\ref{ino}) is rather involved and its integrable structure was not yet fully explored. 
The limit $\kappa=0$ corresponds to the Haldane-Shastry spin chain, whose monodromy matrix was constructed in \cite{BGHP}.  A systematic way
to construct the integrals of motion was given in \cite{HT}. 
Another limit which is tractable  is $L\to \infty $, when the interaction strength becomes proportional to $1/\sinh^2\kappa n$. In this case, the conserved 
quantities
can be obtained \cite{BGHP} from those of the Haldane-Shastry model by the exchange of the real and imaginary periods $L\to i\pi/\kappa$.  The Yangian 
generators are also obtainable by this procedure. The two limiting cases mentioned above, Haldane-Shastry and infinite length Inozemtsev model share the 
particular feature that the Yangian is a symmetry of the Hamiltonian. This can be understood easily from the fact that the conserved quantities are determined \cite{HT} by (a special limit of) the quantum determinant, which commutes with the monodromy matrix.
When $\kappa\to \infty$, we retrieve the conserved quantities of the XXX spin chain. Here, we 
are interested in exploring the first few corrections to the XXX limit. The expansion parameter will be the gauge theory coupling constant $g^2$, related to
$\kappa$ by the relation \cite{SS}
\begin{equation}
\label{kappag}
e^{-2\kappa}=g^2-3g^4+\CO(g^6)\;. 
\end{equation}

 The strategy of the construction is as follow: we use the mondromy matrix of the Haldane-Shastry spin chain and then we continue it 
 analytically in the positions of the spins, $z_j=e^{2\pi ij/L}\to e^{2j\kappa}$. The algebra between the generators is not affected by this change 
 (but the spectrum of the operators will be).  
The key elements which allow to incorporate the long-range nature of the interaction \cite{BGHP} are a set of mutually commuting 
operators known as the Dunkl operators \cite{Dunkl,Polychronakos:1992zk}
\begin{equation}
\label{Dunklcomm}
[d_i,d_j]=0\;.
\end{equation}
In the case of Inozemtsev spin chain, they have the following expression
\begin{equation}
\label{DunklIno}
d_i^I=\sum_{j;j>i}\Theta_{ij}K_{ij}-\sum_{j;j<i}\Theta_{ji}K_{ij}=\sum_{j;j\neq i}\Theta_{ij}K_{ij}-\sum_{j;j<i}K_{ij}
\end{equation}
where $\Theta_{ij}=z_i/(z_i-z_j)$ and $z_j=e^{2j\kappa}$. The operators $K_{ij}$ permute the coordinates $K_{ij}z_j=z_i K_{ij}$.
Under the coordinate permutations, the Dunkl operator behave as
\begin{eqnarray}
\label{Dunklperm}
[K_{i,i+1},d_k]=0\quad {\rm if}\ \ \  k\neq i,i+1\;,\\
K_{i,i+1}d_i-d_{i+1}K_{i,i+1}=1\;. \nonumber
\end{eqnarray}
The Dunkl operators of the Haldane-Shastry and Inozemtsev spin chains are particular in the sense that they are linear in the permutations $K_{ij}$. The fact that they commute is based on the following identity satisfied by the functions $\Theta_{ij}$ defined above
\begin{equation}
\label{Dunklgen}
\Theta_{ij}\Theta_{ik}=\Theta_{ij}\Theta_{jk}+\Theta_{ik}\Theta_{kj}\;
\end{equation}
for any three indices $i,j,k$. If we relax this condition by taking $\Theta_{ij}=1/(1-e(j-i))$, with $e(n)$ some function, the commutation relations (\ref{Dunklcomm}) will not satisfied and we will have to  add to $d_i$'s  higher odd powers
in the permutations to compensate for this defect
\begin{equation}
\label{Dunkhigher}
d_i=d_i^I+\sum_{\{j,k,l\}\neq i} \eta_{\;i;jkl}\,K_{ij}\,K_{ik}\,K_{il}+\CO(K^5)\;.
\end{equation}
The coefficients $\eta_{\;i;jkl}$ will be determined in terms of $\Theta_{ij}$ by the condition of vanishing of the terms quadratic in $K_{ij}$ in the commutators.
We believe that this procedure will allow to determine recursively all the coefficients in the expansion (\ref{Dunkhigher}) and to explicitly build the integrable long-range spin chains with multi-spin interaction we describe below.  

Given the Dunkl operators, one can easily construct the monodromy matrix \cite{BGHP}. The construction in the rest of this section is rather general, and at this point we do not have to specify the precise details of the Dunkl operators, except that they are built from the coordinates 
permutations $K_{ij}$, as explained above. We define  then
\begin{equation}
\label{mmI}
T_a(u)\equiv \bpi (\hat T_a(u))\;, \qquad  \hat T_a(u)=\prod_{j=1}^L \(1+\frac{iP_{ja}}{u-i\,d_j-i/2}\)\;, 
\end{equation}
where $a$ stands for the auxiliary space. The projection operator $\bpi$ acts in the following way: the coordinate permutations $K_{ij}$ are brought to the right
of the expression and then they are replaced with the spin permutations $P_{ij}$. This amounts to working on  (wave)functions which are
symmetric by simultaneous permutations of spins and coordinates, and then freezing the coordinates to some specific values. It was shown in \cite {BGHP} that
$T(u)$ obeys the Yang-Baxter equation with the rational $R$ matrix $R(u)=u+iP$. The proof goes in two steps: first, one checks that the matrix (\ref{mmI}) without the projection obeys the Yang Baxter equation, which is obvious. Second, it can be shown using the commutation relations (\ref{Dunklperm}), see for example eq. (2.24) in reference \cite{BGHP}, that the unprojected monodromy matrix
$\hat T(u)$ preserves the space of function symmetric by simultaneous permutations of spins and coordinates, so that
\begin{equation}
\label{morphism}
\bpi (\hat T_a(u)\hat T_{a'}(v))=\bpi(\hat T_a(u))\bpi(\hat T_{a'}(v))\;.
\end{equation}
In other words, the BGHP projection \cite{BGHP} is a morphism.
In virtue of this relation, the projected matrix obeys the Yang Baxter equation as well.
For the Inozemtsev spin chain, the matrix (\ref{mmI}) is defined when $L\to \infty$, and  is not known how to build the monodromy matrix for a periodic finite chain. However, since we are interested in expanding the Hamiltonian in powers of $g^2$, according to  (\ref{kappag}), on a finite chain we can truncate  the sum (\ref{DunklIno})
to sites from $1$ to $L$. The price to pay is that the resulting Hamiltonian, while remaining integrable, will not be perfectly periodic anymore and some terms connecting the first few and last few sites will be missing. This Hamiltonian will be slightly different from the one considered in \cite{GV}, which probably explains
the difference with the contact terms which appear in that reference.
Here we will concentrate on (long) spin chains with the integral structure coming from (\ref{mmI}), and we hope to elaborate more on the finite chains and periodic boundary conditions elsewhere \cite{Finite}.

 The normalization of the matrix (\ref{mmI}) is chosen such that at $d_j=0$ we retrieve the
standard expression of the monodromy matrix of the XXX model\footnote{The $i$ in front of $d_j$ insures the right ratio 
between $P_{0j}$ and $d_j$, {\it cf.} \cite{BGHP} .} .
From the definition of the projection $\bpi$ it follows that the symmetric combinations of Dunkl operators, for example the symmetric sums,  have the property
\begin{eqnarray}
\label{commproj}
 \bpi \Big( F(d) \sum_i d_i^n\Big)=\bpi \Big( F(d)\Big) \bpi \Big(\sum_i d_i^n\Big)\;,
\end{eqnarray}
where $F(d)$ is an arbitrary function of $d_i$'s.
We can therefore define as mutually commuting quantities the symmetric sums
\begin{eqnarray}
\label{ssd}
\dd_n=\bpi \Big(\sum_i d_i^n\Big)\;, \qquad [\dd_n,\dd_m]=0\;.
\end{eqnarray}
They cannot be used to generate the Hamiltonians of the model, because they have the the same value on all the states \cite{BGHP,HT}. This means that their
eigenvalues can be determined on any state, for example on $|\Omega\rangle\equiv|\uparrow\uparrow\ldots\uparrow\rangle$. 

From the expression (\ref{mmI}) its is straightforward to obtain the Yangian generators by considering the coefficients of the expansion 
around $u=\infty$,
\begin{eqnarray}
\label{yanggen}
&Q_0^{ab}=\sum_{j} E_j^{ab}\;, &\\ \nonumber
\\
&Q_1^{ab}=\sum_{j>i}E_j^{ac}E_i^{cb}+\sum_{j} E_j^{ab}\,\bpi ( d_j)\;, &
\end{eqnarray}
where $E_{j}^{ab}$ are the elementary generators of $gl(2)$ so that $P_{ij}=E_{i}^{ab}E_{j}^{ba}$.

Let us mention here that the construction of the monodromy matrix based on Dunkl operators can be done completely similarly for the $gl(n)$ case. The boundary case works as well, as explained in \cite{bHS}. There, two type of boundary reflection matrices were considered, one preserving the $su(2)$ symmetry, the other which breaks it. 

The monodromy matrix (\ref{mmI}) strongly resembles the monodromy matrix of the inhomogeneous XXX model defined as
\begin{equation}
T_0(u;\theta)= \prod_{j=1}^L \(1+\frac{iP_{ja}}{u-\theta_j-i/2}\) \;.
\end{equation}
It would be tempting to identify $\theta_j$ with the eigenvalue of $d_j$, but this 
cannot be done because of the projection $\bpi$. However, there is a way
to relate the two models.  Let us first use the monodromy matrix acting on the reference state $|\Omega\rangle$. Since the Dunkl operators mutually commute and commute with the spin operators, we can formally expand the matrix $T(u)$
in powers of $d_j$'s. Up to trivial powers of $i$, the coefficients of the expansion are the same as for the expansion of $T_0(u; \theta)$ in powers of $\theta_j$'s,
\begin{eqnarray}
\nonumber
T(u)|\Omega\rangle=\sum_n\sum_{j_1<\ldots<j_n}\sum_{k_1,\ldots,k_n} \frac{i^{k_1+\ldots k_n}}{k_1!\ldots k_n!}\; 
\partial_{j_1}^{\,k_1}\ldots  \partial_{j_n}^{\,k_n} \;T_0(u;\theta)\Big\vert_{\theta=0}    \; \bpi \(d_{j_1}^{\,k_1}\ldots d_{j_n}^{\,k_n}\)|\Omega\rangle
\end{eqnarray}
\vskip-30pt
\begin{eqnarray}
\nonumber &{\rm or}&\\
&T(u)|\Omega\rangle=\CD_\theta \;T_0(u;\theta)|\Omega\rangle \Big\vert_{\theta=0}&
\end{eqnarray}
where the theta operator defined as
\begin{equation}
\label{thetaop}
\CD_\theta=\sum_n\sum_{j_1<\ldots<j_n}\sum_{k_1,\ldots,k_n} \frac{i^{k_1+\ldots k_n}}{k_1!\ldots k_n!}\; C^{k_1,\ldots,k_n}_{j_1,\ldots,j_n}
\;\partial_{j_1}^{\,k_1}\ldots  \partial_{j_n}^{\,k_n} \;,
\end{equation}
\begin{equation}
C^{k_1,\ldots,k_n}_{j_1,\ldots,j_n}\,|\Omega\rangle\equiv \bpi \(d_{j_1}^{\,k_1}\ldots d_{j_n}^{\,k_n}\)\,|\Omega\rangle\;,
\end{equation}
bears strong resemblances to the theta derivative defined in  \cite{GV} at two-loop and three-loop order\footnote{In a previous version of this article,
we have stated that the two differential operators are the same. In fact, we became aware that there are some subtle differences between the present definition and the one 
in \cite{GV}, concerning in particular the terms which are odd in the derivatives, see below. We thank N. Gromov for explaining their definition of the differential operator.}. 
From the property (\ref{commproj}) of the projection we deduce that, if $G (\theta)$ is a symmetric function of $\theta$, we obtain
\footnote{The same property was postulated \cite{KtalkPI} for theta derivative
proposed by N. Gromov and P. Vieira  \cite{GV};  here it is a straightforward consequence of the properties of the Dunkl operators.}
\begin{equation}
\label{dthetasym}
\CD_\theta[ F (\theta) \;G (\theta)]|_{\theta=0}=\CD_\theta[ F (\theta)]|_{\theta=0}\; \CD_{\theta'}[ G (\theta')]|_{\theta'=0}\;.
\end{equation}
The action of the theta operator on any symmetric function of the impurities can be obtained by  working out to the action 
of $\CD_\theta$  on the basis of symmetric functions $p_{n_1}\ldots p_{n_k}$ with  $p_n=\sum_j\theta_j^n$ and $p_0=L$.

Let us define a more general operator ${\mathscr{D}}_\theta$, via
\begin{eqnarray}
\label{thetamor}
T(u)&=&{\mathscr{D}}_\theta \;T_0(u;\theta) \Big\vert_{\theta=0}\\
&\equiv&\sum_n\sum_{j_1<\ldots<j_n}\sum_{k_1,\ldots,k_n} \frac{i^{k_1+\ldots k_n}}{k_1!\ldots k_n!}\; 
\;\partial_{j_1}^{\,k_1}\ldots  \partial_{j_n}^{\,k_n} \;T_0(u;\theta) \Big\vert_{\theta=0}\bpi \(d_{j_1}^{\,k_1}\ldots d_{j_n}^{\,k_n}\)
\nonumber
\end{eqnarray}
so that we can handle the monodromy matrix without acting on a particular state on the right. The coefficients of the differential operator are now  spin operators,
and they cannot be replaced by numbers anymore.
Due to the morphism property (\ref{morphism}) of the projection $\bpi$, we deduce that the operator ${\mathscr{D}}_\theta$ is also a morphism, 
which we call the Bernard-Gaudin-Haldane-Pasquier (BHGP) morphism
\begin{eqnarray}
\label{tBGHPmor}
T_a(u)T_{a'}(v)={\mathscr{D}}_\theta \;T_{0,a}(u;\theta) \Big\vert_{\theta=0} {\mathscr{D}}_\theta \;T_{0,a'}(u;\theta) \Big\vert_{\theta=0}={\mathscr{D}}_\theta \;[T_{0,a}(u;\theta)T_{0,a'}(v;\theta)] \Big\vert_{\theta=0}
\end{eqnarray}
Let us emphasize that the purely differential operator ${\cal{D}}_\theta$ does not possess this property, rather
\begin{eqnarray}
\label{thmor}
{\cal{D}}_\theta \;[T_{0,a}(u;\theta)T_{0,a'}(v;\theta)] \Big\vert_{\theta=0}={\cal{D}}_\theta \;T_{0,a}(u;\theta) \Big\vert_{\theta=0} {\cal{D}}_\theta \;T_{0,a'}(u;\theta) \Big\vert_{\theta=0}+{\rm cross\ terms}\;.
\end{eqnarray}
In (\ref{tBGHPmor}) the cross terms are taken into account by the commutator of $\bpi \(d_{j_1}^{\,k_1}\ldots d_{j_n}^{\,k_n}\)$ with $T_{0,a'}(u;\theta)$.
Upon acting on  the vacuum state we retrieve the original theta operator action
\begin{eqnarray}
\label{vacvac}
{\mathscr{D}}_\theta \;[\ldots](u;\theta) \Big\vert_{\theta=0}|\Omega\rangle={\cal{D}}_\theta \;[\ldots](u;\theta) \Big\vert_{\theta=0}|\Omega\rangle
\end{eqnarray}
where the dots stand for any product of elements of the monodromy matrix.

{\bf Two and three loops from the Inozemtsev model:} 
Let us give an example how to perturbatively compute the operators $d_i^2$ and $d_i\,d_j$ and  their projection on the vacuum state 
$|\Omega\rangle$ using their expression in the Inozemtsev spin chain (\ref{DunklIno}) .
The connection with the dilatation operator of the ${\cal{N}}=4$ theory works up to three loops \cite{SS}, so we are going to expands to this order.
From their definition, it is clear that the coefficients $\Theta_{ij}$ are invariant by translation, $\Theta_{ij}=\Theta_{i+n,j+n}$. 
The projection $\bpi$ amounts to replacing $K_{ij}$ by $P_{ij}$ which evaluates to $1$ on the spin symmetric state $|\Omega\rangle$. Up to terms of the order
$g^6$ we have
\begin{eqnarray}
\bpi(d_i^2)|\Omega\rangle &=&\left[-2g^2+\CO(g^6)\right]|\Omega\rangle\;, \\
\bpi(d_i d_{i+1})|\Omega\rangle &=&\left[g^2-2g^4+\CO(g^6)\right]|\Omega\rangle
\\
\bpi(d_i d_{i+2})|\Omega\rangle &=&\left[2g^4+\CO(g^6)\right]|\Omega\rangle\;,
\end{eqnarray}
while the other combinations quadratic in the $d_i$'s do not contribute at the order we are considering.
Close to the boundaries $i=1,L$ the result of the projection is affected  by the boundary conditions and  is in general different from the bulk 
one. In particular, we have $\bpi(d_i)|\Omega\rangle=0$, except at the boundary. 
 Collecting the (bulk) $g^2$ terms we obtain
\begin{eqnarray}
\label{dtwotwo}
\CD^{ I}_\theta=1+ g^2D_{2,2}+\CO(g^4)\;, \qquad   D_{2,2}\equiv\frac{1}{2}\sum_i\(\partial_{i} -\partial_{i+1}\)^{\,2}\;,
\end{eqnarray}
with $\partial_i= \partial_{\theta_i}$.  Since we are not particularly interested here in specifying the boundary terms, we do not specify the limits of summation above. Let us however mention that these terms are important for insuring the properties of the theta operator defined in (\ref{thetaop}).

We denote by $D_{m,n}$ the term appearing at order $g^m$ and containing $n$th powers of the derivatives. 
The expression (\ref{dtwotwo}) in the bulk coincides with the result of Gromov and Vieira \cite{GV} for the $B(u)$ 
element of the monodromy matrix at two loops.
At order $g^4$ we have some contributions from the terms cubic in the Dunkl operators
\begin{eqnarray}
d_i ^2d_{i+1}=-2g^4+\CO(g^6)&\qquad &d_i ^2d_{i+2}= g^4+\CO(g^6)\\
d_i d_{i+1}^2= 2g^4+\CO(g^6)&\qquad &d_i d_{i+2}^2=-g^4+\CO(g^6)\;,
\end{eqnarray}
where  we should understand these relation after projection and action on the vacuum state (that we omit in the following to avoid cumbersome formulae).
The quadratic and cubic part in the derivatives at order $g^4$  are reproduced by the operator  
\begin{equation}
\!\!D_{4,2}+D_{4,3}=\!\sum_j \left[(\partial_{j}-\partial_{j+2})^2-(\partial_{j}-\partial_{j+1})^2+\frac{i}{6}(\partial_{j}-
\partial_{j+2})^3-\frac{i}{3}(\partial_{j}-\partial_{j+1})^3
\right].
\end{equation}
This contribution seems rather innocuous, since it vanishes on functions which are symmetric in $\theta$. 
 It is absent   from \cite{GV}, where it is  postulated that the odd-order derivatives are absent. Ruling out the odd-order derivatives might still preserve the relation (\ref{dthetasym}), but it will probably introduce unnecessary cross-terms.
 According to N. Gromov, $D_{4,2}$ may be retrieved in their formulation if the Hamiltonian is modified by a transformation which does not change the spectrum.
 
 For the terms quartic in the Dunkl operators, we have two kind of contributions: one is given by two clusters of operators of the type $d_i^2$ or $d_id_{i+1}$
situated in generic position with $i< j$ 
\begin{eqnarray}
d_i ^2d_{j}^2=4g^4+\CO(g^6)\;, &\qquad &d_{i-1} d_{i}d_{j}d_{j+1}=g^4+\CO(g^6)\;,\\
d_i ^2d_{j}d_{j+1}=-2g^4+\CO(g^6)\;,&\qquad & d_{i}d_{i+1}d_j ^2=-2g^4+\CO(g^6)\;,
\end{eqnarray}
and the second type is when the two clusters come close to each other (here we list only the cases which are different from the general case above)
\begin{eqnarray}
d_i ^4=6g^4+\CO(g^6)&\qquad &d_i ^2d_{i+2}^2=3g^4+\CO(g^6)\\
d_i ^2d_{i+1}d_{i+2}=-g^4+\CO(g^6)&\qquad &d_i ^3d_{i+1}=-3g^4+\CO(g^6)\\
d_i d_{i+1}^2d_{i+2}=\CO(g^6)&\qquad &d_id_{i+1}^3=-3g^4+\CO(g^6)\;.
\end{eqnarray}
These contributions can be summarized in
\begin{eqnarray}
D_{4,4}=\frac{1}{8}\sum_{i,j}(\partial_i-\partial_{i+1})^2(\partial_j-\partial_{j+1})^2-\frac{1}{4}\sum_i(\partial_i-\partial_{i+1})^2(\partial_{i+1}-\partial_{i+2})^2
\end{eqnarray}
so that, up to terms of order $g^6$ the operator ${\cal D}^I_\theta$ in the bulk is given by
\begin{eqnarray}
{\cal D}^{I,\,{\rm bulk}}_\theta=1+g^2D_{2,2}+g^4(D_{4,2}+D_{4,3}+D_{4,4})+\CO(g^6)\;.
\end{eqnarray}
To compute the action of $\CD_\theta^I $ on the basis of symmetric sums $p_{n_1}\ldots p_{n_k}$ with $p_n=\sum_j \theta_j^n$ 
it is enough to use the property  (\ref{dthetasym})
\begin{eqnarray}
\label{factD}
\CD_\theta^I \; p_{n_1}\ldots p_{n_k} \vert_{\theta=0}=\prod_{j=1}^k \(\CD_\theta^I\; p_{n_j} \vert_{\theta=0}\)\;.
\end{eqnarray}
By direct computation we get $ \CD_\theta^I \,p_0=L$, $ \CD_\theta^I \, p_2=2g^2L$, $ \CD_\theta^I \, p_4=6g^4L$. When checking the above factorization explicitly we get some subleading $1/L$
corrections in the products, but they are an artifact of working with finite chain and imposing periodic boundary 
conditions $d_{L+1}=d_1$, which are obviously not compatible with the definitions of the Dunkl operators 
(\ref{DunklIno}). As a consequence of (\ref{factD}), on an arbitrary power of the resolvent $G_\theta(u)$ defined as
\begin{eqnarray}
G_\theta(u)=\partial_u \ln \prod_j(u-\theta_j) =\sum_j\frac{1}{u-\theta_j}= \sum_{n\geq 0}\frac{p_n}{u^{n+1}}
\end{eqnarray}
 we have
\begin{eqnarray}
\label{powerk}
\CD_\theta^I \; G^k_\theta(u)\Big\vert_{\theta=0}=\!\!\( {L}\frac{d }{du}\ln f(u,g)\)^k\;.
\end{eqnarray}
This relation implies that the substitution $G_\theta(u) \to {L}\frac{d }{du}\ln f(u,g)$ can be done on any function 
which depends on $\theta$ only through the resolvent
$G_\theta(u)$.
The eigenvalues $a(u)$ and $d(u)$ of the diagonal elements $A(u)$ and $B(u)$ on the vacuum can be similarly evaluated
\begin{eqnarray}
a(u)= f(u^+)^L\;,  \qquad d(u)= f(u^-)^L\;,
\end{eqnarray}
with $u^\pm=u\pm i/2$.
The function $f(u,g)$, determined above has the same expansion to order $g^4$ as the Zhukovsky variable
\begin{eqnarray}
\label{fx}
x(u)=\frac{1}{2}\(u+\sqrt{u^2-4g^2}\)=f(u,g)+\CO(g^6)\;.
\end{eqnarray}
 At the order $g^6$, the dispersion relation for the Inozemtsev model starts to be different of that of 
 the ${\cal N}=4$ dilatation operator, and from that of the BDS model. For the full Inozemtsev model, the all-loop 
 expression of the function $f(u,g(\kappa))$ is given \cite{ino02} by the parametric equation
 \begin{eqnarray}
\label{fxall}
 \frac{f(u^+)}{f(u^-)}=e^{ip}\;, \quad u(p )=\frac{p}{2\pi i \kappa}\zeta_1\left(\frac{i\pi}{2\kappa}
\right)-\frac{1}{2 i \kappa}\zeta_1\left(\frac{i p}{2\kappa}
\right)\;,
\end{eqnarray}
 with the relation between $\kappa$ and $g$ from eq. (\ref{kappag}). Here $\zeta_1(z)$ is the elliptic zeta function with periods $1$ and $i\pi/\kappa$.
 
 We believe that the construction above works for a large class of functions $f(u)$. In \cite{Boost} a procedure was given to construct very general 
 spin chains with long range interaction, and the Inozemtsev model falls in the category of chains constructed from boost  charges in that work. 
 For these models at least, the Dunkl operators and the theta derivatives should exist, and we should have 
 \begin{eqnarray}
\dd_n \, |\Omega\rangle=(-i)^n L C_n \,|\Omega\rangle
\end{eqnarray}
with $C_n$ the coefficients of the expansion 
\begin{eqnarray}
\frac{d }{du}\ln f(u)=\frac{1}{u}\sum_{n\geq 0}\frac{C_n}{u^n}\;.
\end{eqnarray}
We hope it is possible to prove this formula by giving an explicit construction of the  Dunkl operators (\ref{Dunkhigher}).  

{\bf The BDS spin chain:} The BDS spin chain, defined through its Bethe ansatz \cite{BDS}, 
 \begin{eqnarray}
\label{bdsa}
\left[\frac{x(u_k^+)}{x(u_k^-)}\right]^L=\prod_{j\neq k}^M\frac{u_k-u_j+i}{u_k-u_j-i}\;,
\end{eqnarray}
was devised such as to reproduce the all-loop dispersion relation
of the dilatation operator in the $su(2)$ sector.
When supplemented with the BES dressing phase \cite{BES} it reproduces the $su(2)$ dilatation operator 
at all loop. 
It was noticed in \cite{BDS} that the BDS ansatz can be formally derived from the inhomogeneous XXX ansatz with inhomogeneities given by
$\theta_j=2g \sin 2\pi j/L$. 
In \cite{RSS} it was shown that the BDS spin chain is in fact a projection of the one-dimensional Hubbard model 
at half filling to the spin sector. In that construction, $\theta_j=2g \sin q_j\sim i d_j$ are in fact dynamical variables corresponding to the momenta $q_j$ of the underlying fermions, 
and in principle subject to backreaction from the spin degrees of freedom. 
The action of the 
 theta operator  at all-loop is given by
\begin{eqnarray}
\label{thetaconj}
\; \CD^\BDS_\theta\; p_{n}\Big\vert_{\theta=0}=i^L{\bf p}_n=L C^\BDS_{n} \;
\end{eqnarray}
with 
\begin{eqnarray}
C^\BDS_{n}=(2g)^n \int_{-\pi}^\pi \frac{dq}{2\pi} \sin^{n} q\;,\!\quad \! {\Rightarrow}\!  \quad \! C^\BDS_{2n}=\frac{(2g)^{2n}\,
\Gamma(n+1/2)}{\sqrt{\pi}\Gamma(n+1)}\;, \! \quad\!  C^\BDS_{2n+1}=0\;.
\end{eqnarray}
The backreaction decouples from the computation of the symmetric sums, as it can be seen in appendix E of \cite{RSS}.
Now, in virtue of the equation (\ref{powerk}) we  can substitute for any $F[G_\theta(u)]$
\begin{eqnarray}
\CD^\BDS_\theta \; F[G_\theta(u)]\Big\vert_{\theta=0}=F\left[\frac{L }{\sqrt{u^2-4g^2}}\right]  \;,
\end{eqnarray}
which is of course consistent with the BDS equations (\ref{bdsa}).

{\bf Arbitrary number of magnons:} The relations above were given for one-magnon eigenstates, 
but generalization to arbitrary magnon eigenstates is straightforward, due to the morphism property (\ref{tBGHPmor}). Let us consider the
product of $M$ copies of the monodromy matrix with auxiliary spaces $a_1,\ldots, a_M$. Since the 
unprojected monodromy matrix preserves the space of functions which are symmetric under 
simultaneous permutations of coordinates and spins \cite{BGHP}, the product of projections is equal to the projection of the product.
Proceeding exactly as before, we get that
\begin{eqnarray}
\label{multipleTD}
T_{a_1}(u_1)\ldots T_{a_M}(u_M) |\Omega\rangle={\cal D}_\theta \ T_{a_1,0}(u_1;\theta)\ldots T_{a_M,0}(u_M;\theta)\; |\Omega\rangle\Big\vert_{\theta=0}\;.
\end{eqnarray}
From this expression, one can extract any monomial in the operators $A(u)$,  $B(u)$, $C(u)$, $D(u)$ by appropriate projections in the auxiliary spaces  $a_i$.
In particular, we have 
\begin{eqnarray}
\label{bbb}
B(u_1)\ldots B(u_M)|\Omega\rangle={\cal D}_\theta\; B_0(u_1;\theta)\ldots B_0(u_M;\theta)\,\;|\Omega\rangle\Big\vert_{\theta=0}&\equiv&  |\{u\}\rangle_g \;,\\
\langle\Omega|C(v_1)\ldots C(v_M)=\langle\Omega|\,{\mathscr D}_\theta\;C(v_1;\theta)\ldots C(v_M;\theta)\Big\vert_{\theta=0}&\equiv&\,_g\langle\{v\}|\;.
\label{ccc}
\end{eqnarray}
Due to the fact that the $R$ matrix in the Yang-Baxter equation is the same as in the XXX model, the algebra of the $A(u)$,  $B(u)$, $C(u)$, $D(u)$
operators is the same\footnote{See for example \cite{faddeev}.}, with the only modification that the 
eigenvalues of $A(u)$ and $D(u)$ on the vacuum $|\Omega\rangle$ are now $a(u)=f(u^+)^L$ and $b(u)=f(u^-)^L$. 
We conclude that the vectors defined in (\ref{bbb}) are eigenvectors of the BDS Hamiltonian if the rapidities $\{u\}$ satisfy the BDS ansatz equations (\ref{bdsa}).

{\bf Scalar products and the correlation functions:} 
In \cite{Tailor1} the setup was given to compute the correlation functions of the single-trace operators of the ${\cal N}=4$ gauge theory in the weak coupling limit.
The building blocks entering the correlation functions are the Slavnov-type scalar products \cite{Slavnov}
\begin{eqnarray}
\CS_{\{u;\theta\},\{v;\theta\}}=\langle \{v;\theta\}|\{u;\theta\}\rangle 
\end{eqnarray}
where $|\{u;\theta\}\rangle \equiv B_0(u_1;\theta)\ldots B_0(u_M;\theta)\,|\Omega\rangle$ and, for example, the rapidities $\{u\}$ obey Bethe ansatz equations with impurities $\theta$. The variables $\{v\}$ may obey a different Bethe ansatz. The tree-level correlation functions can be obtained \cite{Tailor1} by setting $\theta=0$. The scalar 
products are given by the Slavnov determinant formula \cite{Slavnov,Wheeler,Foda}.
As observed by direct computation up to two loop order in \cite{GV} and shown above, at higher loop order the Bethe ansatz  eigenstates  are deformed according to
(\ref{bbb})
 into 
\begin{eqnarray}
|\{u\}\rangle\ \  \rightarrow\ \  |\{u\}\rangle_g=\CD_\theta |\{u;\theta\}\rangle\Big\vert_{\theta=0}
\end{eqnarray}
and their scalar products can be evaluated using (\ref{multipleTD}), or equivalently (\ref{bbb},\ref{ccc}), as 
\begin{eqnarray}
\nonumber
_g\langle \{v\}|\{u\}\rangle_g&=&\CD_\theta\,\langle\Omega| \,C_0(v_1;\theta)\ldots C_0(v_M;\theta)B_0(u_1;\theta)\ldots B_0(u_M;\theta)\,|\Omega\rangle\Big\vert_{\theta=0}\\
&=&\CD_\theta\;\CS_{\{u;\theta\},\{v;\theta\}}\Big\vert_{\theta=0}\;. 
\end{eqnarray} 
Let us emphasize that, due to the morphism property (\ref{tBGHPmor}) of $\mathscr{D}_\theta$ and to the property (\ref{vacvac}), there are no cross terms in the scalar product, as there are in \cite{GV}. This is an important difference between our definitions and theirs, and it may give a considerable computational advantage at higher loop.
The Slavnov determinant  depends on the impurities $\theta$  only via the resolvents $G_\theta(u)$, $G_\theta(v)$. If the theta operator
is given by the all-loop expression (\ref{thetaconj}),  the only effect of acting with it on the Slavnov determinant  is to substitute
\begin{eqnarray}
\label{subsBDS}
G_\theta(u) \longrightarrow L\frac{d\ln f(u)}{du}\;
\end{eqnarray}
and to insure that $\{u\}$ (and possibly $\{v\}$) obey the BDS ansatz (\ref{bdsa}).
This would work even on expressions with finite number of magnons, up 
to $1/L$ terms\footnote{A careful treatment of the boundary terms for the periodic system and of the  Hamiltonian insertion 
at the splitting point can be found in \cite{GV}. In this reference it was found at two loops, or one loop for the correlator, that the two corrections compensate each other. 
  }. 
In fact, to compute the scalar products we do not really need to know about the theta operator; it is enough to know that the monodromy matrix exists and that
its matrix elements obey the same algebra as for XXX, with modified functions $a(u)$ and $d(u)$. This is insured by the morphism properties
of the BGHP projection. Having an 
explicit value for the Dunkl operators allows to determine the microscopic realization of the model ({\it i.e.} to determine the Hamiltonians).  

The quasiclassical limit of the Slavnov determinants, when the distribution of mag-nons $\{u\}$ and $\{v\}$  
condense on some contours $\Gamma_{\bf u}$ and $\Gamma_{\bf v}$ was computed very recently by I. Kostov \cite{Ivan3pt}, by using a fermionic representation.  
The result, which generalizes previous results \cite{Tailor1,Tailor23} is
\begin{eqnarray}
 \la{intexplnreg} \ln \CS_{\{u\},\{v\}}&=&\oint \limits_ {C_{\bf v}} \frac{dz}{2\pi } \ \text{Li}_2\left(
e^{i(G_{\bf u}(z) + G_{\bf v}(z)- G_\theta (z)}\) - \oint \limits_{C_{\bf u} }
 \frac{dz}{2\pi } \ \text{Li}_2\left( e^{ i G_ {\bf v}(z) - iG_{\bf u}(z)}\) \no
 \\
& &{\rm with} \quad G_{{\bf u},{\bf v}}(z)=\int_{\Gamma_{{\bf u},{\bf v}}} \!\!\!dz' \;\frac{\rho_{{\bf u},{\bf v}}(z')}{z-z'} 
\end{eqnarray}
and the closed contours $C_{{\bf u},{\bf v}}$ encircling $\Gamma_{{\bf u},{\bf v}}$ counterclockwise. We refer to \cite{Ivan3pt} for a discussion of the subtleties in closing the contour
$C_{\bf u}$ around the singularities of the dilogarithm. Here we have omitted for simplicity some terms coming from the normalization of the states and which can be reconstituted by comparison with \cite{Ivan3pt}. These terms will drop out anyway from the physically meaningful quantities which are the scalar products divided by the norms.
The results at tree level is obtained by setting all the impurities to zero. 
It is straightforward to generalize this result at all loop for the BDS model, as well as for the other similar long-range models, via the substitution (\ref{subsBDS}). 
In these cases, we define the quasi momentum $p(u)$  through
\begin{eqnarray}
\label{qmBDS}
e^{2ip_{\bf u}(u)}=\left[\frac{f(u^-)}{f(u^+)}\right]^L\prod_{j}^M\frac{u-u_j+i}{u-u_j-i}\;.
\end{eqnarray}
In the classical limit $u\sim L$, so $p_{\bf u}(u)$ is related to the resolvent via
\begin{eqnarray}
\label{cqmBDS}
p_{\bf u}(z)= G_{\bf u}(z)-\frac{L}{2}\frac{d\ln f(z)}{dz}\;.
\end{eqnarray}
and the Bethe ansatz equations (\ref{bdsa}) become in this limit 
\begin{eqnarray}
\label{ccqmBDS}
p\!\!\!/(z)=\pi n\;, \quad {\rm or}\quad G\!\!\!\!/_{\bf u}(z)=\pi n +\frac{L}{2}\frac{d\ln f(z)}{dz}\;, \quad {\rm for}\quad z\in \Gamma_{\bf u}\;.
\end{eqnarray}
Therefore, the results of \cite{Ivan3pt} can be expressed uniquely in terms of the quasi momentum and the potential ${d\ln f(z)}/{dz}$.
The norm eigenvectors  of the Hubbard model was conjectured by G\"ohman and Korepin \cite{GK}; it would be interesting to compare the 
the result at half filling with the prediction for the BDS case. 

{\bf Consequence for the correlation functions for AdS/CFT.}
The all-loop Bethe ansatz for the dilatation operator in the $su(2)$ sector can be obtained from (\ref{bdsa}) by supplementing it with the 
BES dressing phase $\sigma_\BES$ \cite{BES}. In terms of quasi momentum, this amounts to substituting 
\begin{eqnarray}
\label{qqmBES}
p(u)\longrightarrow p^\BES(u)=p^\BDS(u)-i\ln \sigma^\BES(u)\;.
\end{eqnarray}
and in the continuum limit 
\begin{eqnarray}
\label{qqqmBES}
p^\BES(z)= G^\BES_{\bf u}(z)-\frac{L}{2\sqrt{z^2-4g^2}}\;.
\end{eqnarray}
When $g$ is large the quasi momentum $p^\BES(z)$ reproduces the algebraic curve data \cite{KMMZ}.
It is remarkable that (\ref{intexplnreg}) contains only  the quasi momentum $p(z)$ and the potential $x'(z)=1/\sqrt{z^2-4g^2}$ via 
$G_\theta(z)$. The simplicity of this result leads us, see also \cite{Ivan3pt}, to conjecture that the dressing phase can be incorporated in the results of 
\cite{Ivan3pt}
simply by substituting the resolvents 
\begin{eqnarray}
\label{subsBES}
G_{{\bf u},{\bf v}}(z) \longrightarrow G^\BES_{{\bf u},{\bf v}}(z)\;.
\end{eqnarray}
According to this conjecture, and assuming that the cancelation between the contact terms and the Hamiltonian insertions observed in \cite{GV} survives at higher loop, the all-loop three-point correlators in the $su(2)$ sector would be given by
\begin{eqnarray}
 &&\ln \la{clasC123} { C_{123}(g) } =-\hf \!
\sum_{j=u,v,w}\;\oint \limits_{ C_j} \frac{dz}{2\pi } \ \Li \big[e^{ 2i p_j(z)}\big]
 \\
&&
\!\!\! + \oint   \limits_{ \tilde C^\infty_\uu\cup C_{v}}
\!\!\!\!\!\!\!\!   \frac{dz}{2\pi }  
\ \text{Li}_2\big[e^{  i p_{{\bf u}}(z) + i p_{\bf v}(z)+i{L_3}/{2\sqrt{z^2-4g^2}}}
\big]  + \oint \limits_{C_{\bf w}} \frac{dz}{2\pi } 
  \text{Li}_2\left[ e^{ i { (L_2-L_1)/2\sqrt{z^2-4g^2} } + i p_{\bf w}(z) } \right] 
\no
\end{eqnarray}
with the quasimometa $p_j(z)$ defined as
\begin{eqnarray}
 p_j(z)=G^\BES_j(z)-\frac{L_j}{2\sqrt{z^2-4g^2}}\;, \quad{\rm and}\quad ({\bf u},{\bf v},{\bf w})\equiv (1,2,3)\;.
\end{eqnarray}
A general strategy to compute the higher loop correlators was also suggested in the conclusions of  \cite{Tailor1},  where the authors noticed that the substitution of the full quasi momentum does not work  for finite size systems. Here we conjecture that (\ref{clasC123}) is true in the quasi classical limit, when at least the length  $L_1$ is large and the corresponding rapidities are large, $u_i\sim L_1$, and condense on some cuts.
If the quasi classical limit of the full solution of the three-point function depends only on the quasi momenta, then it should be relatively easy to generalize the  
result to correlators of three operators in arbitrary positions by using the algebraic curve data \cite{KMMZ, BeisertK}. The recent results \cite{Janik}
about the Wilson loops and correlators of a Wilson loop and a local operator also seem to point out in this direction.
The way the quasi momentum appears in the classical limit of the Slavnov products reminds the thermodynamical Bethe 
ansatz for the spectrum \cite{Gromov09}. It would be instructive to clarify the interrelation between the correlators, the amplitudes/Wilson lines 
and the TBA for the spectrum.

{\bf Acknowledgements:} The author thanks N. Gromov,  O. Foda, I. Kostov, V. Pasquier, P. Vieira  and D. Volin for illuminating discussions,  D. Volin  for writing a  Mathematica code for the Dunkl operators algebra, and N. Gromov and P. Vieira for sharing their updated result for the three-loop theta derivative \cite{GV} for comparison,  for explaining their exact definition of the theta  morphism and for their constructive criticism on the previous versions of the preprint.


\begin{thebibliography}{99}

\bibitem{AdSCFT}
J.~M. Maldacena,
 Adv. Theor. Math. Phys. {\bf 2}, 231 (1998), {\tt hep-th/9711200},
A.~M. Polyakov, 
 Int. J. Mod. Phys. {\bf A14}, 645 (1999), {\tt hep-th/9809057}, 
E.~Witten, Adv.Theor.Math.Phys.{\bf 2}:253-291,(1998).

\bibitem{habilitation}
D.~Serban, 
{\it ``Integrability and the AdS/CFT correspondence''}, 
J.Phys. A {\bf 44}:124001 (2011).
{\tt arXiv:1003.4214}

\bibitem{Beisert-Rev}
N.~{Beisert {\it et al}}, ``{Review of AdS/CFT Integrability: An Overview},''
  {Letters in Mathematical Physics} {\bf 99} (Jan., 2012) 3--32,
{{\tt arXiv:1012.3982}};


\bibitem{RV}
  R.~Roiban and A.~Volovich,
  {\it ``Yang-Mills correlation functions from integrable spin chains,''}
  JHEP {\bf 0409} (2004) 032
 {\tt  [hep-th/0407140]}

\bibitem{Tailor1}
J. Escobedo, N. Gromov, A. Sever and P. Vieira, 
{\it ``Tailoring three-point functions and integrability''}, 
{\tt arXiv:1012.2475}

\bibitem{Tailor23}
J. Escobedo, N. Gromov, A. Sever and P. Vieira, 
{\it ``Tailoring Three-Point Functions and Integrability II. Weak/strong coupling match''}, {\tt arXiv:1104.5501};
N.~{Gromov}, A.~{Sever}, and P.~{Vieira},
 {\it ``Tailoring Three-Point Functions
  and Integrability III. Classical Tunneling''}
   {{\tt arXiv:1111.2349}}.
   
 \bibitem{Orselli}  
 A. Bissi, T. Harmark, M. Orselli
{\it   ``Holographic 3-point function at one loop''}, {\tt arXiv:1112.5075}.

   \bibitem{GV}
N. Gromov, P. Vieira,
{\it Quantum Integrability for the Three-Point Functions},
{\tt arXiv:1202.4103}


\bibitem{aTBA}
L.~F.~Alday, D.~Gaiotto and J.~Maldacena, {\it ``Thermodynamic Bubble Ansatz,''} JHEP 1109 (2011) 032, {\tt arXiv:0911.4708 };
L.~F.~Alday, J.~Maldacena, A.~Sever and P.~Vieira, {\it `` Y-system for Scattering Amplitudes,''} J. Phys. A A 43 (2010) 485401, {\tt arXiv:1002.2459}

\bibitem{wlTBA}
N. Drukker, {\it ``Integrable Wilson loops,''} {\tt arXiv:1203.1617};
D. Correa, J. Maldacena and A. Sever, {\it ``The quark anti-quark potential and the cusp anomalous dimension from a TBA equation,''}{\tt arXiv: 1203.1913}

\bibitem{Janik}
R.~A.~Janik, P.~Laskos-Grabowski,
{\it ``Surprises in the AdS algebraic curve constructions - Wilson loops and correlation functions''}, {\tt arXix:1203.4246}.

\bibitem{Okuyama:2004bd} 
  K.~Okuyama and L.~-S.~Tseng,
  {\it ``Three-point functions in N = 4 SYM theory at one-loop,''}
  JHEP {\bf 0408}, 055 (2004)
  {\tt hep-th/0404190.}

\bibitem{Izergin}
A.~G. {Izergin},
 ``{\it Partition function of the six-vertex model in a finite
  volume},'' 
  { Soviet Phys. Doklady} {\bf 32} (1987) 878.

\bibitem{Korepin}
V.~E. Korepin,
{\it``Calculation of norms of Bethe wave functions,'' }
 {
  CMP} {\bf 86} (1982) 391-418.
  
  \bibitem{Slavnov}
N.~A. Slavnov, 
{\it ``The algebraic Bethe ansatz and quantum integrable systems,''}
  { Russian Mathematical Surveys} {\bf 62} (2007), no.~4, 727.
  
  \bibitem{Foda}
O.~{Foda}, 
``{\it N=4 SYM structure constants as determinants},'' 
{{\tt arXiv:
  1111.4663}}.
  
    \bibitem{Ivan3pt}
I. Kostov, {\it ``Classical Limit of the Three-Point Function from Integrability''}, to appear.

\bibitem{BKS}
N.~Beisert, C.~Kristjansen and M.~Staudacher,
{\it ``The dilatation operator of N = 4 super Yang-Mills theory,''}
Nucl.\ Phys.\ B {\bf 664} (2003) 131,
{\tt hep-th/0303060}.

\bibitem{SS}
Serban D., Staudacher M.
 {\it ``Planar ${\cal N}=4$ gauge theory and the Inozemtsev long range spin chain''}, 
JHEP 001 (2004). 

\bibitem{ino02}
V.~I.~Inozemtsev,
{\it ``Integrable Heisenberg-van Vleck chains with variable range exchange,''}
Phys.\ Part.\ Nucl.\  {\bf 34} (2003) 166,
{\tt hep-th/0201001}. 

\bibitem{BDS}
N.~Beisert, V.~Dippel, and M.~Staudacher, 
``{\it A novel long range spin chain and
  planar N = 4 super Yang- Mills},''
   { JHEP} {\bf 07} (2004) 075,
{{\tt hep-th/0405001}}.

\bibitem{Dunkl}
C.~F.~Dunkl,
{\it``Differential-Difference Operators Associated to Reflection Groups''},
Trans.  Am. Math. Soc.
Vol. 311, {\bf 1} (1989), 167.

\bibitem{Polychronakos:1992zk} 
  A.~P.~Polychronakos,
  {\it ``Exchange operator formalism for integrable systems of particles,'' }
  Phys.\ Rev.\ Lett.\  {\bf 69}, 703 (1992)
{\tt hep-th/9202057}.

  \bibitem{RSS}
   A.~ Rej, D.~ Serban, M.~Staudacher
   {\it``Planar N=4 Gauge Theory and the Hubbard Model''}
 JHEP 0603:018, (2006), {\tt hep-th/0512177}.
   
 \bibitem{Boost}
T.~Bargheer, N~Beisert, F.~Loebbert,
  {\it ``Boosting Nearest-Neighbour to Long-Range Integrable Spin Chains''}
J.Stat.Mech, 0811:L11001, (2008).


\bibitem{Polychronakos:1992ki} 
  A.~P.~Polychronakos,
  {\it ``Lattice integrable systems of Haldane-Shastry type,''}
  Phys.\ Rev.\ Lett.\  {\bf 70}, 2329 (1993)
 {\tt  hep-th/9210109}.
 
 
  \bibitem{BGHP}D.~Bernard, M.~Gaudin, F.~D.~M.~Haldane 
and V. ~Pasquier, 
{\it  ``Yang-Baxter equation in long-range interacting systems,''} 
J.Phys.A: Math.Gen {\bf 2 6} (1993) 5219,
{\tt hep-th/9301084}.

\bibitem{HT} J.~C.~Talstra, F.~D.~M.~Haldane,
{\it ``Integrals of motion of the Haldane Shastry Model,''} J. Phys. A {\bf 28}
(1995) 2369, 
{\tt cond-mat/9411065}.

\bibitem{KtalkPI}
N. Gromov, Talk at the Perimeter Institute, March 1012,
{\tt http://pirsa.org/displayFlash.php?id=12020162}.

 \bibitem{bHS} 
D. Bernard, V. Pasquier et D. Serban,
{\it Exact Solution of Long-Range Interacting Spin Chains with Boundaries}, 
 Europhysics Letters, {\bf 30} (5),
1995 pgs 301-306.

  \bibitem{Finite} D. Serban, in progress.
   
  \bibitem{faddeev}
  L.D. Faddeev
   {\it ``How Algebraic Bethe Ansatz works for integrable model''}, {\\tt hep-th/9605187}.
 
  \bibitem{Wheeler}
M.~{Wheeler},
``{\it An Izergin-Korepin procedure for calculating scalar products
 in the six-vertex model},''
   { Nucl. Phys. B} {\bf 852} (2011)
  468, {{\tt arXiv:1104.2113}}.

\bibitem{GK}
F. G\"ohman, V. E. Korepin,
{\it ``The Hubbard chain: Lieb-Wu equations and norm of the eigenfunctions''}
Phys.Lett. {\bf A263} (1999) 293-298
{\tt cond-mat/9908114}

\bibitem{BES}
N.~Beisert, B.~Eden, and M.~Staudacher, 
``{\it Transcendentality and crossing},''
  { J. Stat. Mech.} {\bf 0701} (2007) P021,
{{\tt hep-th/0610251}}.

\bibitem{KMMZ}
V. Kazakov, A. Marshakov, J. Minahan, K. Zarembo, {\it ``Classical/ quantum integrability in AdS/CFT,''} JHEP 0405 (2004) 024, {\tt hep-th/0402207}

\bibitem{BeisertK}
N. Beisert, V. Kazakov, K. Sakai, K. Zarembo, {\it ``The Algebraic Curve of Classical 
Superstrings on AdS5 x S5,''} Commun. Math. Phys. 263 (2006) 659, {\tt hep-th/0502226}

 \bibitem{Gromov09}
N.~Gromov
{\it ``Y-system and Quasi-Classical Strings''}
 JHEP 1001:112 (2010), {{\tt arXiv:0910.3608}}.
 

\end{thebibliography}
\end{document}